\documentclass{article}
\usepackage{graphicx}

\newcommand{\bdel}{\mbox{\boldmath{$\Delta$}}}
\newcommand{\be}{\begin{equation}}
\newcommand{\ee}{\end{equation}}

\begin{document}

\begin{center}
{\bf UNIFIED BCS-LIKE MODEL OF PAIRING AND ALPHA-CORRELATIONS}\vspace{0.5cm}

Roman SEN'KOV$^{1}$ and Vladimir ZELEVINSKY$^{2}$
\end{center}
\begin{center}
\noindent{\small $^{1}$ {\sl Physics Department, Central Michigan
University, Mount Pleasant, MI 48859, USA\\
$^{2}$ Department of Physics and Astronomy and National Superconducting
Cyclotron Laboratory, Michigan State University, East Lansing, MI 48824-1321,
USA }}
\end{center}

\begin{abstract}

We construct a BCS-like model that combines nucleonic pairing correlations
and possible quartic correlations of alpha-type in a single variational
wave function and derive corresponding gap equations. In the approximation of
large logarithms typical for the BCS approach, we show that the system
reveals two possible types of a condensate which cannot coexist.
If the alpha-type condensate prevails at $N=Z$, the growth of the neutron
excess will naturally lead to the first order phase transition to the
nucleon condensates.

\end{abstract}

\section{Introduction}

Half a century ago it was understood that pairing interaction playing an
extremely important role in the structure of nuclei is qualitatively similar to
the interaction between the electrons in superconducting metals \cite{bohr58}.
This analogy very soon lead to intensive development of microscopic theory
of nuclear pairing \cite{belyaev59,migdal59,soloviev61,thouless62} that was
shown to have also significant new elements due to the smallness of the system
and discrete single-particle spectrum. Enormous experimental information
clearly shows that pairing not only explains the odd-even effects and the structure
of low-lying excitations; pairing considerably changes vibrational and rotational
motion, level densities, cross sections of many reactions, regularities of
nuclear decays and fission etc. The very existence of many exotic nuclei is
possible only due to the extra binding that comes from pairing. The recent
development of mesoscopic physics revealed that similar pairing effects are
important in the systems like atomic clusters, organic molecules, Fullerenes,
and cooled atoms in traps \cite{delft01,kresin08,kresinlittle90,deaven95,jeremy09};
here the theory can be built using nuclear physics as a prototype.

There exists an approximate conservation law in atomic nuclei that has no
counterpart in the electron system, namely the isospin. The isospin invariance
is a distinctive symmetry of strong forces, with generalizations to higher
symmetries in the quark world. This approximate symmetry is violated by Coulomb
forces \cite{auerbach83} and by electromagnetic effects in small mass
differences of the light quarks and, as a consequence, neutron-proton mass
difference. In low-energy nuclear physics the isospin invariance is a good
guideline and, as such, is widely used in many versions of the shell model
\cite{brown88,brown01}. Assuming that isospin $T$ is a good quantum number of a
nuclear stationary state, we should consider various pairing modes, with $T=1$
and $T=0$ of a pair \cite{lane64,goodman79}.

Quasideuteron $n-p$ ($T=0$) pairs are possible only for the specific values of
the angular momentum $J$ of the pair, that is not allowed for identical
particles, for example $J=0$ is forbidden. In a finite system, such as the nucleus,
the condensate of pairs with $J\neq 0$ and therefore the trend to
``ferromagnetic" alignment, is energetically less favorable compared to
the condensate of spherically symmetric pairs with $J=0$. In spite of the
existence of $T=0$ correlations, especially in exotic nuclei along the $N=Z$ line,
which are currently under scrutiny with radioactive beam facilities, most probably
such a condensate is not present in the ground states. Different estimates
\cite{VZ03,bertsch09} indicate that the isoscalar pairing could develop in
the ground state of a medium or heavy nucleus if the corresponding interaction
strength would be stronger than in reality by a factor $\sim (2.5-3)$.

Systems with $N\approx Z$ can amplify another type of correlations, namely
quartic correlations of the alpha-type. With a large number of particles, we
cannot expect a considerable probability of the genuine bosonic
alpha-condensate that was hypothesized for light nuclei \cite{tohsaki01,belyaeva09},
such as $^{12}$C (famous Hoyle state) or $^{16}$O. However, the presence of usual
pairing with the large coherence length may coexist with quartic correlations
creating four-nucleon clusters of large size. In standard alpha-decay and
various reactions with knock-out of alpha-particles \cite{sakharuk97,sakharuk99}, the
process may start with clusters of a different structure adjusted to the
existing shell occupancies \cite{janouch83}.
The clusters proceed through restructuring on the way to the
continuum. Apart from the reactions, such correlations could give rise to
smearing of the shell structure and collectivity persistent in the regions of
nominally closed shells, as it is observed in quadrupole transitions near
$^{100}$Sn, the last doubly magic $N=Z$ nucleus \cite{starosta07}.

The idea of coexistence, mutual support and/or competition of the usual pairing
and alpha-correlations, was repeatedly discussed in the literature, for example
\cite{apostol87,dumitrescu90,koh96,baran06}. A remote analog of this problem
can be found in the consideration of the BCS-BEC crossover for cold fermionic
atoms \cite{giorgini08}. If, with the growing pairing strength, one observes the
crossover from the paired phase with the large coherence length to the gas of
strongly bound molecules, then the residual interaction between the molecules
can lead to the formation of quartic entities (dimers), or even polymers. A similar
(metastable) situation is related to biexcitons in semiconductors and exciton condensation \cite{eisenstein04}; electron-hole bilayer quantum dots share some pairing features
with complex atoms and nuclei \cite{steffi04}. In the present paper we generalize
the problem of $T=1$ pairing solving it
together with the quartic interaction of alpha-type. We find the analytic
BCS-like solution that, to the best of our knowledge, was not presented earlier
in the literature.

\section{Formulation of the problem}

\subsection{Hamiltonian}

We consider a system of $N$ neutrons and $Z$ protons in a certain time-reversal
invariant mean field that generates single-particle orbitals $|\nu\tau)$, where
$\tau=\pm 1/2$ is the isospin projection, and $\nu$ combines all other quantum
numbers. According to the Kramers theorem, the orbitals come in degenerate
time-conjugate pairs, $|\nu)$ and $|\tilde{\nu})$, and for the fermions the double
time-reversal works as $|\tilde{\tilde{\nu}})=-|\nu)$. Assuming that the
pairing forces prefer the pairs of time-conjugate single-particle states, we
use the pair operators ($T=1; t=T_{3}=0,\pm 1$)
\begin{equation}
A_{Tt}(\nu)=\sum_{\tau\tau'}C^{Tt}_{1/2\tau\,1/2\tau'}a_{\nu\tau}a_{\tilde{\nu}\tau'},
\quad A^{\dagger}_{Tt}(\nu)=\sum_{\tau\tau'}C^{Tt}_{1/2\tau\,1/2\tau'}
a^{\dagger}_{\tilde{\nu}\tau'}a^{\dagger}_{\nu\tau},   \label{1}
\end{equation}
\begin{equation}
B^{\dagger}_{Tt}(\nu)=\sum_{\tau\tau'}(-)^{1/2-\tau'}C^{Tt}_{1/2\tau\,1/2\tau'}
a^{\dagger}_{\nu \tau}a_{\nu-\tau'}.                             \label{2}
\end{equation}
For each value of $\nu$, the six pair operators $A_{1t},A^{\dagger}_{1t}$,
the number operator $\sim B_{00}$ and the isospin operator $\sim B_{1t}$
form a set of ten generators of the ${\cal SO}(5)$ algebra \cite{hecht65}
generalizing the ${\cal SU}(2)$ algebra \cite{kerman61} used in the pairing
problem for one type of fermions.

The Hamiltonian of the problem will include the mean field part,
\begin{equation}
H_{1}=\sum_{\nu\tau}\epsilon_{\nu\tau}a^{\dagger}_{\nu\tau}a_{\nu\tau},
\label{3}
\end{equation}
where single-particle energies are counted from the appropriate chemical potentials
$\mu_{\tau}$; the isovector pairing interaction between all kinds of particles,
\begin{equation}
H_{2}=-\,\sum_{t t', \nu, \nu'>0}G_{\nu t,\nu' t'}
A^{\dagger}_{1t}(\nu) A_{1t'}(\nu'),                           \label{4}
\end{equation}
where $\nu>0$ means that the pair of single-particle orbits, $\nu$ and $\tilde{\nu}$, is counted only once,
and the additional quartic interaction \cite{dumitrescu90},
\begin{equation}
H_{4}=-\,\sum_{\nu, \nu'>0}C_{\nu\nu'}\alpha^{\dagger}_{\nu}\alpha_{\nu'},
                                                                     \label{5}
\end{equation}
where we introduced the creation, $\alpha^{\dagger}_{\nu}$, and annihilation,
$\alpha_{\nu}$, operators for the alpha-like clusters living in a cell $\nu$
that includes two proton and two neutron occupying $\nu$ and $\tilde{\nu}$ orbits,

\begin{equation}
\alpha^{\dagger}_{\nu}=A^{\dagger}_{11}(\nu)A^{\dagger}_{1-1}(\nu).       \label{6}
\end{equation}

\subsection{ Ground state wave function}

The natural generalization of the BCS variational approach is given by the trial
wave function of the ground state,
\begin{equation}
|\Psi_{0}\rangle=\prod_{\nu>0}\Bigl(u_{\nu}+\sum_{t=0,\pm 1} v_{\nu t}A^{\dagger}_{1t}(\nu)
+z_{\nu}\alpha^{\dagger}_{\nu}\Bigr)|0\rangle,                        \label{7}
\end{equation}
Here we included all possibilities for pairwise time-conjugate occupation of the cell $\nu$,
namely the absence of particles ($u_{\nu}$), one pair with isospin projection $t$
($v_{\nu t}$) and the full four-cluster ($z_{\nu}$). Such an extended ansatz was used
earlier \cite{chasman03} as an input to numerical calculations and without alpha-interaction.
Introducing the isovector ${\bf v}_{\nu}$ with the ``length" $|{\bf v}_{\nu}|^{2}=\sum_{t}|v_{\nu t}|^{2}$,
we normalize the wave function (\ref{7}) according to
\begin{equation}
\prod_{\nu>0}\Bigl(|u_{\nu}|^{2}+|{\bf v}_{\nu}|^{2}+|z_{\nu}|^{2}\Bigr)=1. \label{8}
\end{equation}

We can
redefine the five complex constants which determine the wave function
(\ref{7}) in each cell $\nu$ in terms of the proton-neutron language as ($v_{\nu 0}$ does not change)
\begin{equation}
u_{\nu}=u_{\nu n}u_{\nu p}, \quad v_{\nu 1}=u_{\nu p}v_{\nu n}, \quad v_{\nu -1}=
u_{\nu n}v_{\nu p}, \quad z_{\nu}=v_{\nu n}v_{\nu p},             \label{9}
\end{equation}
where we set $\tau_{n}=1/2,\,\tau_{p}=-1/2$. Then the trial wave function can be
presented in the form
\begin{equation}
|\Psi_{0}\rangle=\prod_{\nu>0}\left[v_{\nu 0}A^{\dagger}_{10}(\nu)+\Bigl(u_{\nu n}+
v_{\nu n}A^{\dagger}_{11}(\nu)\Bigr)\Bigl(u_{\nu p}+v_{\nu p}A^{\dagger}_{1-1}(\nu)\Bigr)
\right]|0\rangle,                                               \label{10}
\end{equation}
with the normalization condition
\begin{equation}
|v_{\nu 0}|^{2}+\Bigl(|u_{\nu n}|^{2}+|v_{\nu n}|^{2}\Bigr)
\Bigl(|u_{\nu p}|^{2}+|v_{\nu p}|^{2}\Bigr)=1.                 \label{11}
\end{equation}
We can conclude that the wave function factorizes into independent proton and
neutron parts in formal analogy to the standard BCS-type pairing if $v_{\nu 0}$
vanishes. However, such a conclusion would be premature since (i) the
occupation parameters of the two parts are mutually dependent through the
interaction and (ii) this does not work in the case of the pure
alpha-condensate when the only non-zero coefficients in the trial function
(\ref{7}) are $u_{\nu}$ and $z_{\nu}$, while the pairs are absent, and ${\bf
v}=0$.

The expectation values of the number operators for the wave function
(\ref{7}) are given by
\begin{equation}
N=\sum_{\nu>0}\Bigl(2|v_{\nu 1}|^{2}+|v_{\nu 0}|^{2}+2|z_{\nu}|^{2}\Bigr),
\quad Z=\sum_{\nu>0}\Bigl(2|v_{\nu -1}|^{2}+|v_{\nu 0}|^{2}+
2|z_{\nu}|^{2}\Bigr),                            \label{12}
\end{equation}
while the total number of nucleons in this state is
\begin{equation}
A=N+Z=\sum_{\nu>0}\Bigl(2|{\bf v}_{\nu}|^{2}+4|z_{\nu}|^{2}\Bigr). \label{13}
\end{equation}
The expectation values of various terms of the Hamiltonian in the state (\ref{7})
are found straightforwardly: for the one-body term, where we introduce
$\epsilon_{\nu 1}=\epsilon_{\nu n},\,\epsilon_{\nu -1}=\epsilon_{\nu p}$, and
$\epsilon_{\nu 0}=(\epsilon_{\nu n}+\epsilon_{\nu p})/2$, and assume that
the chemical potentials are included in the definition,
\begin{equation}
\langle\Psi_{0}|H_{1}|\Psi_{0}\rangle=\sum_{\nu>0}\Bigl(\sum_{t}2\epsilon_{\nu
t}|v_{\nu t}|^{2}+4\epsilon_{\nu 0}|z_{\nu}|^{2}\Bigr);       \label{14}
\end{equation}
for the pairing term
\begin{equation}
\langle\Psi_{0}|H_{2}|\Psi_{0}\rangle=-\,\sum_{tt',\nu,\nu'>0}
G_{\nu t,\nu't'}\Bigl(u_{\nu}v^{\ast}_{\nu t}-(-)^{t}z^{\ast}_{\nu}v_{\nu
-t}\Bigr)\Bigl(u^{\ast}_{\nu'}v_{\nu' t'}-(-)^{t'}z_{\nu}v^{\ast}_{\nu' -t'}\Bigr),
                                                          \label{15}
\end{equation}
where we, as it is usually assumed in the BCS approximation, neglected the
terms with $\nu=\nu'$ which give the correction to one-body energies being
smaller than the pairing terms with $\nu\neq \nu'$ by a factor $\propto
\Omega$, the effective volume of the single-particle space; and the quartic
term,
\begin{equation}
\langle\Psi_{0}|H_{4}|\Psi_{0}\rangle=-\,\sum_{\nu,\nu'>0}C_{\nu\nu'}
u_{\nu}z^{\ast}_{\nu}u^{\ast}_{\nu'}z_{\nu'},                   \label{16}
\end{equation}
where we again neglect the contributions $\nu=\nu'$.

\subsection{Variational equations}

Now we try to minimize the ground state energy with respect to the amplitudes
of the trial wave function (\ref{7}). It is convenient to account for the
normalization condition (\ref{8}) assuming that the factors in eq. (\ref{8})
are equal to one in each cell. Then we can always consider $u_{\nu}$ to be real
and express its variation with respect to $v^{\ast}_{\nu t}$ and
$z^{\ast}_{\nu}$ as
\begin{equation}
\delta u_{\nu}=-\,\frac{1}{2u_{\nu}}\Bigl(\sum_{t}v_{\nu t}\delta v^{\ast}_{\nu
t}+ z_{\nu}\delta z^{\ast}_{\nu}\Bigr).                             \label{17}
\end{equation}
The analog of the BCS energy gap carries now the isospin index and emerges in the form
\begin{equation}
\Delta_{\nu t}=\sum_{t',\nu'>0}G_{\nu t,\nu't'}\Bigl(u_{\nu'}v_{\nu't'}-(-)^{t'}
z_{\nu'}v^{\ast}_{\nu'-t'}\Bigr).                                   \label{18}
\end{equation}
In the presence of the quartic interaction, a new quantity of the same type
appears in the process of variation,
\begin{equation}
\Delta^{(\alpha)}_{\nu}=\sum_{\nu'>0}C_{\nu\nu'}u_{\nu'}z_{\nu'}.
                                                                  \label{19}
\end{equation}
Collecting all terms of the Hamiltonian, taking the expectation values and
performing variations, we obtain two equations which can be written as
\begin{equation}
v_{\nu t}=\,\frac{u_{\nu}\Delta_{\nu t}-(-)^{t}z_{\nu}\Delta^{\ast}_{\nu -t}}
{2\epsilon_{\nu t}+B_{\nu}
}, \label{20}
\end{equation}
and
\begin{equation}
z_{\nu}=\,\frac{-\sum_{t}(-)^{t}v_{\nu -t}\Delta_{\nu t}+u_{\nu}
\Delta^{(\alpha)}_{\nu}}{4\epsilon_{\nu 0}+B_{\nu}
},                                    \label{21}
\end{equation}
where the additional notations are introduced
\begin{equation}
B_{\nu}=\,\frac{1}{2u_{\nu}}\Bigl[\sum_{t}v^{\ast}_{\nu t}\Delta_{\nu t} +
z^{\ast}_{\nu}\Delta^{(\alpha)}_{\nu}
\Bigr]+({\rm c.c.}) \equiv \,
\frac{1}{u_{\nu}}\,{\rm Re}({\bf v}^{\ast}_{\nu}\cdot
\bdel_{\nu}+z^{\ast}_{\nu}\Delta^{(\alpha)}_{\nu}).     \label{22}
\end{equation}
Here (and in many equations below and above) we use boldface vectors in isospin space.
In fact, the parameter $B_{\nu}$ can be introduced as a Lagrange multiplier corresponding
to the normalization condition (\ref{8}), when instead of using the condition (\ref{17})
we can look
for a minimum of $\langle\Psi_{0}|H|\Psi_{0}\rangle + B \langle\Psi_{0}|\Psi_{0}\rangle$.

\subsection{Blocking energy}

In our model the Hamiltonian $H=H_{1}+H_{2}+H_{4}$ contains kinetic (one-body),
pairing (two-body), and alpha (four-body, or quartic) terms. Then the expectation
value of ground state energy with the amplitudes satisfying variational equations
is given by the sum of
\begin{equation}
\langle\Psi_{0}|H_{1}|\Psi_{0}\rangle=\sum_{\nu>0}\Bigl\{2 \left[ u_{\nu}^{2}B_{\nu}
+|z_{\nu}|^{2}(B_{\nu}+4\epsilon_{\nu 0}) \right]-B_{\nu} - 2 u_\nu z_\nu
 \Delta^{(\alpha)}_{\nu}\Bigr\},                       \label{23}
\end{equation}
\begin{equation}
\langle\Psi_{0}|H_{2}|\Psi_{0}\rangle=-\sum_{\nu>0} \Bigl\{u_{\nu}^{2}B_{\nu}
+|z_{\nu}|^{2}(B_{\nu}+4\epsilon_{\nu 0}) - 2 u_\nu z_\nu  \Delta^{(\alpha)}_{\nu}\Bigr\},
                                                     \label{24}
\end{equation}
and
\begin{equation}
\langle\Psi_{0}|H_{4}|\Psi_{0}\rangle=-\sum_{\nu>0} u_\nu z_\nu  \Delta^{(\alpha)}_{\nu}.
                                              \label{25}
\end{equation}
The total energy, therefore, equals
\begin{equation}
E_{0}=\sum_{\nu>0}\Bigl\{u_{\nu}^{2}B_{\nu}+|z_{\nu}|^{2}(B_{\nu}+4\epsilon_{\nu 0})
- u_\nu z_\nu  \Delta^{(\alpha)}_{\nu} - B_{\nu}\Bigr\}.         \label{26}
\end{equation}
One can block the 4-cell inserting a pair proton+neutron with the same $\nu$
(this ``wrong" pair occupies $|\nu_{p})$ and $|\nu_{n})$ single-particle orbits)
and therefore
eliminating all pairing processes that include this cell as an initial or a final state.
In this way we would lose the contributions of $(H_{1}+2H_{2}+2H_4)_{\nu}=-B_{\nu}$ with
the subscript $\nu$ and come to an excited state with excitation energy
\begin{equation}
E_{\nu}= B_{\nu}+{\cal E}_{\nu},                         \label{27}
\end{equation}
where ${\cal E}_{\nu}$ is an energy of an object inserted in the cell instead of the
proper condensate pair, for example a mean-field energy $\epsilon_{\nu n}+\epsilon_{\nu p}=
2\epsilon_{\nu 0}$ of a ``wrong" pair. In the usual BCS case we would have for this
blocking energy
\begin{equation}
E_{\nu}=2\sqrt{\epsilon_{\nu}^{2}+\Delta_{\nu}^{2}}\equiv 2e_{\nu} \quad  \to \quad
B_{\nu}= 2e_{\nu}-2\epsilon_{\nu 0}.                         \label{28}
\end{equation}
In our more general case, blocking energy has the same physical meaning.

\subsection{Symmetry considerations}

Here we consider the amplitudes of the ground state wave function in a given 4-cell $\nu$
and omit the subscript $\nu$. It is also possible to choose all amplitudes real.
In terms of isospin vectors, the variational equations in this cell take the form
\begin{equation}
{\bf v}=\hat{E}^{-1}(u{\bdel}+z{\bdel}'), \quad z=\,\frac{({\bdel}'\cdot{\bf v}) + u
\Delta^{(\alpha)}}
{E+2\epsilon_{0}}, \quad u=\,\frac{({\bdel}\cdot{\bf v}) + z
\Delta^{(\alpha)}}{E-2\epsilon_{0}},                      \label{29}
\end{equation}
where we use the diagonal in isospin space matrix
\begin{equation}
\hat{E}=\,{\rm diag}\,(E+\delta\epsilon,E,E-\delta\epsilon), \quad
\delta\epsilon=\epsilon_{n}-\epsilon_{p}, \quad E=B+2\epsilon_{0},    \label{30}
\end{equation}
and the isospin-vector ${\bdel}'=(\Delta_{p},-\Delta_{0},\Delta_{n})$
with the components $\Delta'_{t}=(-)^{1-t}\Delta_{-t}$, an analog of the time-reversal
operation in isospin space.

If the pairing interaction $H_{2}$ is isospin-invariant so that the matrix elements
$G_{\mu t,\nu t'}=G_{\mu\nu} \delta_{t t'}$ do not depend on the isospin projections, the ansatz
(\ref{7}) is invariant with respect to the generalized particle-hole ($p-h$)
transformation. If in the wave function (\ref{7}) we populate our cells starting
with the empty state $|0\rangle$, we can use as a new reference state the fully
occupied ``Dirac sea", namely the fictitious state $|\infty\rangle$ where all
single-particle positions are filled in by infinitely many particles. Annihilating pairs
and clusters from $|\infty\rangle$ we construct, in every 4-cell, the {\sl p-h conjugate} states
\begin{equation}
|\psi'\rangle=\Bigl\{u+({\bf v}\cdot{\bf A})+z\alpha\Bigr\}|\infty\rangle,  \label{31}
\end{equation}
that can be rewritten as a usual state of type (\ref{7}),
\begin{equation}
|\psi'\rangle=\Bigl\{u\alpha^{\dagger}+({\bf v}'\cdot{\bf A}^{\dagger})+z\Bigr\}|0\rangle, \label{32}
\end{equation}
generated from the original state (\ref{7}) by the transformation in space of the variational
parameters, $u\leftrightarrow z$, and
\begin{equation}
{\bf v}\leftrightarrow {\bf v}'=(v_{p},-v_{0},v_{n})\equiv \hat{E}'^{-1}(u\bdel'+z\bdel), \label{33}
\end{equation}
where $\hat{E}'$ differs from $\hat{E}$, eq. (\ref{30}), by the sign of $\delta\epsilon$ (all
signs of single-particle energies, $\epsilon_{n},\epsilon_{p}$ and $\epsilon_{0}$, are to be inverted).

The contributions of isospin-invariant pairing,
\begin{equation}
\langle H_{2}\rangle= -\sum_{\mu,\nu>0}G_{\mu\nu}\Bigl((u{\bf v}+z{\bf v}')_{\mu}
\cdot(u{\bf v}+z{\bf v}')_{\nu}\Bigr),                         \label{34}
\end{equation}
and quartic interaction,
\begin{equation}
\langle H_{4} \rangle=-\,\sum_{\mu,\nu>0}C_{\mu\nu} (u z)_{\mu}\cdot(u z)_{\nu},
                                                           \label{35}
\end{equation}
are clearly invariant under the {\sl p-h} conjugation. Then the pairing gaps,
\begin{equation}
{\bdel}_{\mu}=\sum_{\nu>0}G_{\mu\nu}(u{\bf v}+z{\bf v}')_{\nu}, \label{36}
\end{equation}
and $\Delta^{(\alpha)}_{\mu}$, eq. (\ref{19}), are invariant too.
To make the one-body contribution $\langle H_{1}\rangle$ invariant we have to change
all signs of single-particle energies, in this case $\langle H_{1}\rangle$ is
shifted only by a constant so that the variational equations do not change. Using this
we can show the invariance of the excitation energies (\ref{27}).

There is an interesting relation between the amplitudes $u, \bf v$, and $z$, that
deserves to be mentioned here. In the absence of alpha-alpha interaction, the amplitudes are connected as follows:
\be
2 u z = 2 v_p v_n - v^2_0.
\ee
This relation does not depend on a particular form of the pairing interaction or on the single-particle part of the Hamiltonian, but assumes that the amplitudes satisfy
the variational equations (\ref{29}).

\subsection{Excitation energies and variational parameters}

If we eliminate the vector parameter ${\bf v}$ in eq. (\ref{29}) and introduce the scalar
product of isospin vectors with the metric defined by the matrix $\hat{E}^{-1}$, eq. (\ref{30}),
\begin{equation}
S({\bf a},{\bf b})\equiv \left({\bf a}\,\frac{1}{\hat{E}}\,{\bf b}\right), \label{37}
\end{equation}
we can present the remaining equations for the parameters $u$ and $z$ in a given cell $\nu$ as
\begin{equation}
(E+2\epsilon_{0})z=S({\bdel}',{\bdel})u+S({\bdel}',{\bdel}')z + \Delta^{(\alpha)}u,
                                                                     \label{38}
\end{equation}
\be
(E-2\epsilon_{0})u=S({\bdel},{\bdel})u+S({\bdel},{\bdel}')z + \Delta^{(\alpha)}z.
                                                                    \label{39}
\ee
Equivalent equations can be derived from the operator equations of motion
with the BCS-type linearization.
The solvability condition determines the excitation energy $E$,
\be
D(E;\epsilon_0,\delta\epsilon,\bdel,\Delta^{(\alpha)})=0,        \label{40}
\ee
where the determinant $D$ equals
\begin{equation}
D=\Bigl[E+2\epsilon_{0}-S({\bdel}',{\bdel}')\Bigr]\Bigl[E-2\epsilon_{0}-S({\bdel},{\bdel})\Bigr]
-\Bigl(S({\bdel},{\bdel}')+\Delta^{(\alpha)}\Bigr)^2                          \label{41}
\end{equation}

This equation for excitation energy $E$ is quite complicated since all $S$-structures
also depend on $E$. Nevertheless, we can write a formal solution for the amplitudes
$u,z$ of a given cell $\nu$ in the following form:
\be
z^{2}=X\,\frac{E-2\epsilon_{0}-S({\bdel},{\bdel})}{2E}, \quad
u^{2}=X\,\frac{E+2\epsilon_{0}-S({\bdel}',{\bdel}')}{2E},           \label{42}
\ee
\be
u z =X\,\frac{\Delta^{(\alpha)}+S({\bdel}',{\bdel})}{2E},         \label{43}
\ee
where $X$ is a normalization constant that can be expressed through the partial derivative
of the determinant $D$ with respect to excitation energy $E$ taken at the root of the secular
equation (\ref{40}) when all other parameters are kept constant,
\be
X=2E \Bigl(\frac{\partial D}{\partial E}\Bigr)^{-1}.              \label{44}
\ee
This relation that can be directly checked follows from the general properties of the Green's
function $(E-H)^{-1}$ that has poles corresponding to the excitation energies of the system
and therefore coinciding with the roots of $D$; the residues at the poles are derived from the
expansion of the determinant near the poles and, at the same time, they determine the normalization condition (\ref{8}).

\section{Gap equations: General form}

The gap equations make the whole set of equations complete and self-consistent.
According to the formal solution (\ref{42}-\ref{44}), the variational amplitudes
for the ground state wave function depend on the pairing gaps that, in turn, depend
on amplitudes. Eliminating ${\bf v}$ and ${\bf v}'$ with the aid of eqs. (\ref{29})
and (\ref{33}), we can present the definitions of the gaps
(\ref{18})
in the following form:
$$ {\bdel}_{\mu}=\sum_{\nu>0}G_{\mu\nu}(u{\bf v}+z{\bf v}')_{\nu}= $$
\be =\sum_{\nu>0}G_{\mu\nu}\left\{\left(u^2\frac{1}{\hat{E}}
+ z^2\frac{1}{\hat{E}'}\right)\bdel
+ uz\Bigl(\frac{1}{\hat{E}}+\frac{1}{\hat{E}'}\Bigr)\bdel'
\right\}_{\nu},                                             \label{45}
\ee
where amplitudes $u^2, z^2, uz$ should be taken from
eqs. (\ref{42}-\ref{44}),
and eq. (\ref{19}) for $\Delta_{\mu}^{(\alpha)}$ has to be added. For any given set
of single-particle levels these equations can be solved numerically
as a function
of interaction parameters. To get a general idea of the character of emerging physics
we will analyze the set of equations in the BCS spirit introducing integrals over
single-particle level density \cite{belyaev59,migdal59} and looking for logarithmic
singularities.

Below we analyze the gap equations substituting the matrix elements $G_{\mu \nu}$ and
$C_{\mu \nu}$ by the effective isospin-invariant constant $G$ and alpha-clustering
constant $C$ within a layer $\pm\Theta$ around the Fermi surface and zero outside
this layer. In nuclei the thickness of the interaction layer should be of the order
of the typical oscillator shell frequency. In such an approximation, all pairing gaps,
$\Delta_{n,p,0}$ and $\Delta^{(\alpha)}$, are constants inside the layer.
With $\rho(\epsilon)\approx \rho_{0}$ as the
density of pair states near the Fermi
surface, we transform the finite sums in eqs. (\ref{45}) and (\ref{19}) to the integrals
over energy within the layer. In this transformation we can take $\epsilon_{\nu 0}$ as
an independent variable $\epsilon$ assuming that in every cell actual single-particle
energies $\epsilon_{n}$ and $\epsilon_{p}$ are defined as functions of this variable.
For example, the gap equation for $\Delta^{(\alpha)}$ comes to the following form:
\be \Delta^{(\alpha)}= C \rho_0 \int
\frac{\Delta^{(\alpha)}+S(\bdel',\bdel)}{2 E} X d\epsilon.    \label{46}
\ee
Next, it is convenient to change the integration variable from $\epsilon$ to energy $E$,
introducing effectively the density of physical excitations. Then, remembering the relation
(\ref{44}) we can rewrite the $X d\epsilon$ as
\be 2E \Bigl(\frac{\partial D}{\partial E}\Bigr)^{-1} d \epsilon =
-2E \Bigl(\frac{\partial D}{\partial \epsilon}\Bigr)^{-1} d E,   \label{47}
\ee
where we assumed that $E$ is a function of $\epsilon$ defined by eq. (\ref{40})
and $\delta\epsilon$ defined in eq. (\ref{30}) is a constant shift, essentially
due to the Coulomb interaction (and difference of chemical potentials for $N\neq Z$).
The partial derivative ${\partial D}/{\partial \epsilon}$ can be directly calculated,
and we finally come to
\be X d \epsilon = \pm \frac{E d E}{2 \sqrt{\Bigl(E-\frac{1}{2}(S(\bdel',\bdel')+S(\bdel,\bdel))\Bigr)^2-
\Bigl(\Delta^{(\alpha)}+S(\bdel',\bdel)\Bigr)^2}}.             \label{48}
\ee
The integral (\ref{46}) and similar integrals for the pairing gaps are well defined but
they are quite complicated since all the $S-$structures are $E$-dependent.

Now we notice that all the gap integrals are logarithmically diverging at high
excitation energy $E$. In fact, this singularity is responsible for the Cooper phenomena
of instability of a normal Fermi surface in the presence of an attractive pairing interaction.
As in the standard BCS theory, we introduced the energy
cut-off parameter $\Theta$ to regularize this divergence assuming that the magnitudes
of the pairing gaps are much smaller than the value of this parameter and we are
interested only in the large logarithmic contributions,
\be \bdel, \Delta^{(\alpha)} \ll \Theta \;\;\; \mbox{and} \;\;\; \ln
\frac{\Theta}{\Delta} \gg 1.                            \label{49}
\ee
The large logarithmic integral that appears in all gap equations is (assuming
the symmetry around the Fermi surface)
\be I=\int^{\Theta} \frac{d E}{\sqrt{\Bigl(E-\frac{1}{2}(S(\bdel',\bdel')
+S(\bdel,\bdel))\Bigr)^2-\Bigl(\Delta^{(\alpha)}+S(\bdel',\bdel)\Bigr)^2}}. \label {50}
\ee
If we neglect all non-logarithmic terms we come up to the following set of equations:
\begin{equation}
\bdel = G \rho_0 I \bdel,                           \label{51}
\end{equation}
\begin{equation}
\Delta^{(\alpha)} = \frac{1}{2} C \rho_0 I  \Delta^{(\alpha)}.     \label{52}
\end{equation}

The immediate important conclusion is that, with logarithmic accuracy, the two
competing condensates with
non-zero energy gaps can appear only in the situation
``either" - ``or", so that the nucleon and alpha-condensates are not compatible.
Indeed, the diverging integral $I$ is the same in both equations (\ref{51}-\ref{52})
while the interactions strengths can coincide only accidentally. Certainly,
the non-condensate correlations of these two types can (and have to) coexist.

\section{Pure pairing solution}

\subsection{Excitation energies and variational parameters}

Let us first go back to the exact equations and look for the solution with $\Delta^{(\alpha)}=0$,
when the excitation energy $E$ satisfies a simplified equation,
\begin{equation}
\Bigl[E+2\epsilon_{0}-S({\bdel}',{\bdel}')\Bigr]\Bigl[E-2\epsilon_{0}-S({\bdel},{\bdel})\Bigr]
=S^{2}({\bdel},{\bdel}').                                   \label{53}
\end{equation}
The positive roots are given by
\begin{equation}
E_{\pm}=\sqrt{\epsilon_{n}^{2}+\epsilon_{p}^{2}+{\bdel}^{2}\pm \sqrt{Q}},
                                                                \label{54}
\end{equation}
where the notation is used
\begin{equation}
Q=\Delta_{0}^{4}+4\Delta_{0}^{2}(\epsilon_{n}\epsilon_{p}-\Delta_{n}\Delta_{p})
+4e_{n}^{2}e_{p}^{2}.                                         \label{55}
\end{equation}
We notice simple meaning of the limiting values of energies (\ref{54}) at
$\Delta_{0}\rightarrow 0$, when they correspond to the creation of a proton-neutron
pair of quasiparticles or to the inversion of the isospin of a quasiparticle,
\begin{equation}
E_{+}\rightarrow e_{n}+e_{p}, \quad E_{-}\rightarrow |e_{n}-e_{p}|; \label{56}
\end{equation}
the standard BCS energies $e_{n,p}$ were defined in eq. (\ref{28}). For each root, we find
the variational parameters (\ref{42},\ref{43}) for a given cell $\nu$ in
terms of corresponding energies (\ref{53}), and
\begin{equation}
X=\,\frac{E^{2}-(\delta\epsilon)^{2}}{E_{+}^{2}-E_{-}^{2}}.     \label{57}
\end{equation}

\subsection{Pairing gap equations}

With isospin-independent pairing matrix elements $G_{\mu\nu}$ and for a given type
of excitations, $E_{+}$ or $E_{-}$, the set of coupled equations takes the form
\begin{equation}
\Delta_{n\mu}=\sum_{\nu>0}G_{\mu\nu}\left(\frac{(E^{2}-e_{n}^{2}+e_{p}^{2})\Delta_{n}
-\Delta_{0}^{2}(\Delta_{n}+\Delta_{p})}{E(E_{+}^{2}-E_{-}^{2})}\right)_{\nu},
                                                               \label{58}
\end{equation}
\begin{equation}
\Delta_{p\mu}=\sum_{\nu>0}G_{\mu\nu}\left(\frac{(E^{2}-e_{p}^{2}+e_{n}^{2})\Delta_{p}
-\Delta_{0}^{2}(\Delta_{n}+\Delta_{p})}{E(E_{+}^{2}-E_{-}^{2})}\right)_{\nu},
                                                               \label{59}
\end{equation}
\begin{equation}
\Delta_{0\mu}=\sum_{\nu>0}G_{\mu\nu}\left(\frac{[E^{2}-(e_{n}-e_{p})^{2}-
(\Delta_{n}+\Delta_{p})^{2}]\Delta_{0}}{E(E_{+}^{2}-E_{-}^{2})}\right)_{\nu}.
                                                               \label{60}
\end{equation}

In the logarithmic approximation, all pairing gaps, $\Delta_{n,p,0}$, are constant
inside the layer. In contrast to the standard BCS case, we encounter here
four types of integrals:
\begin{equation}
I_{a}=G\int d\epsilon\,\rho(\epsilon)\,\frac{f_{a}}{2E\sqrt{Q}}, \quad a=0,1,2,3,  \label{61}
\end{equation}
where
\begin{equation}
f_{0}=1,\; \;f_{1}=E^{2}, \;\;f_{2}=e_{n}^{2}-e_{p}^{2},
\;\;f_{3}=(\epsilon_{n}-\epsilon_{p})^{2}.         \label{62}
\end{equation}
The integral $I_{2}$ is antisymmetric with respect to the isospin mirror reflection,
$n\leftrightarrow p$. The final set of coupled non-linear equations reads
\begin{equation}
\Delta_{n}=(I_{1}-I_{2})\Delta_{n}-I_{0}(\Delta_{n}+\Delta_{p})\Delta_{0}^{2}, \label{63}
\end{equation}
\begin{equation}
\Delta_{p}=(I_{1}+I_{2})\Delta_{p}-I_{0}(\Delta_{n}+\Delta_{p})\Delta_{0}^{2}, \label{64}
\end{equation}
\begin{equation}
\Delta_{0}=(I_{1}-I_{3})\Delta_{0}-I_{0}(\Delta_{n}+\Delta_{p})^{2}\Delta_{0}. \label{65}
\end{equation}
We can again slightly simplify the problem assuming that the possible difference of neutron
and proton chemical potentials in a system with close values of $N$ and $Z$ comes mainly
from a constant Coulomb shift, so that we set $\delta\epsilon=\epsilon_{p}-\epsilon_{n}=$ const.
In this case
\begin{equation}
I_{3}=I_{0}(\delta\epsilon)^{2}.                                              \label{66}
\end{equation}

\subsection{Solutions}

In the trivial case of $\Delta_{0}=0$, the amplitude $v_{0}$ vanishes, and we return to
the simple case of separate neutron and proton BCS pairing. Therefore we look for a
non-trivial solution with all $\Delta_{0,n,p}\neq 0$. Eqs. (\ref{65}) and (\ref{66}) give
\begin{equation}
I_{1}=1+I_{0}[(\Delta_{n}+\Delta_{p})^{2}+(\delta\epsilon)^{2}].          \label{67}
\end{equation}
The combination of eqs. (\ref{63}) and (\ref{64}) provides another relation,
\begin{equation}
2\Delta_{n}\Delta_{p}(I_{1}-1)=I_{0}\Delta_{0}^{2}(\Delta_{n}+\Delta_{p})^{2}. \label{68}
\end{equation}
Thus, the non-zero pairing gaps are interrelated through
\begin{equation}
2\Delta_{n}\Delta_{p}[(\Delta_{n}+\Delta_{p})^{2}+(\delta\epsilon)^{2}]=\Delta_{0}^{2}
(\Delta_{n}+\Delta_{p})^{2}.                                                \label{69}
\end{equation}

Two classes of solutions are evident in the fully symmetric case, $\delta\epsilon=0$,
that corresponds to $N=Z$,
\begin{equation}
\underline{A}: \quad 2\Delta_{n}\Delta_{p}=\Delta_{0}^{2}; \quad \underline{B}:
\Delta_{n}+\Delta_{p}=0.                                           \label{70}
\end{equation}
In the case {\sl A}, we have $({\bdel}\cdot{\bdel}')=0$ and, along with this, $uz=0$. This strange
condition leads to the 4-cells occupied either by pairs and quartets (zero probability of
an empty cell) or without any quartets; both situations are unphysical in a fully symmetric
system. In addition, the physical root of excitation energy (\ref{54}) reduces to
$E_{+}=\epsilon+\sqrt{\epsilon^{2}+{\bdel}^{2}}$, which gives the pairing gap by a factor
of 2 smaller than the alternative solution {\sl B}, $\Delta_{p}=-\Delta_{n}$. The negative sign
does not bring any difficulty since the phase of $\Delta$ is essentially arbitrary
and could be redefined to the opposite from the very beginning. In the case {\sl B} we have
\begin{equation}
1=G\int \,\frac{\rho\,d\epsilon}{2e}, \quad e_{n}=e_{p}\equiv e=\sqrt{\epsilon^{2}+
{\bdel}^{2}/2},                                                  \label{71}
\end{equation}
and the appropriate root for the excitation energy is $E_{+}=2e$. In the solution of type {\sl A},
we would have
\begin{equation}
1=G\int \,\frac{\rho\,d\epsilon}{2\sqrt{\epsilon^{2}+{\bdel}^{2}}}, \label{72}
\end{equation}
with a smaller gap, as it is mentioned above.

In the case of $\delta \epsilon=0$, the system is symmetric with respect to isospin
rotations. We just need to keep the relation $N=Z$ to be fulfilled in average. One
can always find a rotation that brings the $p-n$ amplitude to zero, $v_0=0$.
There is a degeneracy with respect to such rotations that can be lifted by
any extra interaction violating isospin symmetry and
defining the actual structure of the ground state. It can be shown that in the case
of $\delta \epsilon \ne 0$ there is only one solution that belongs to the
type $A$. As seen before, this type of solution is energetically unfavorable and
the ground state will be presented by separate proton and neutron condensates with
$\Delta_0=0$.

\section{Pure alpha-condensate}

The solution of the previous section corresponds to a pure pairing superfluid mode.
Being proportional to the product $uz$, the gap $\Delta^{(\alpha)}$ can be exactly
zero only if $uz=0$ or there is no quartic interaction, $C=0$. Starting with the pairing
solution, the alpha-correlations are defined by the non-logarithmic
terms omitted in eq. (\ref{51}) which would define $\Delta^{(\alpha)}$ as a small
complement to the pairing gaps. Another solution starts with a pure alpha-condensate.
For this case, all pairing gaps $\bdel$ vanish along with the pairing amplitudes ${\bf v}$
(strictly speaking this is possible only for $N=Z$).
We have again to stress here that the statements of this type are only valid in the
asymptotic ``logarithmic" regime when $I \gg 1$.

The simple solution of eq. (52) is well known. With vanishing pairing amplitudes,
${\bf v}=0$, and pairing gapes, $\bdel=0$, all equations are greatly simplified and
can be easily solved analytically. The equation for the excitation energy becomes
\begin{equation}
\Bigl[E+2\epsilon_{0}\Bigr]\Bigl[E-2\epsilon_{0}\Bigr]=(\Delta^{(\alpha)})^2, \label{73}
\end{equation}
with the solutions
\begin{equation}
E =\pm \sqrt{4\epsilon_{0}^2+(\Delta^{(\alpha)})^2}.               \label{74}
\end{equation}
The variational amplitudes are given by
\begin{equation}
z^{2}=\frac{E-2\epsilon_{0}}{2E}, \quad u^{2}=\frac{E+2\epsilon_{0}}{2E},
\quad u z =\frac{\Delta^{(\alpha)}}{2E}.                \label{75}
\end{equation}
Finally, the gap equation takes the form
\begin{equation}
1 = \frac{1}{2} C \rho_0 \int_{\Delta^{(\alpha)}}^{\Theta} \frac{dE}{E} = \frac{1}{2} C \rho_0 \ln \Bigl(\frac{\Theta}{\Delta^{(\alpha)}}\Bigr),         \label{76}
\end{equation}
with the standard solution for the gap
\begin{equation}
\Delta^{(\alpha)} = \Theta \exp \left( - \frac{2}{C\rho_0} \right). \label{77}
\end{equation}

\section{Phase transition}

As we have found, a developed alpha condensate does not coexist with a pair condensate.
Let us assume that, in a nuclear system with $N=Z$, the alpha-alpha interaction is
strong enough to make the alpha-clustering favorable for the ground state.
In this case the ground state energy in the BCS-like approximation can be written as
\begin{equation}
E_{{\rm g.s.}}=E_0 - \frac{1}{2}\,\rho_0 \Delta_{\alpha}^2,    \label{78}
\end{equation}
where $E_0$ is the ground state energy without any pairing or quartic correlations.

For a system with $N \neq Z$, while the difference $|N-Z|$ is not very large,
the alpha condensate still will be the ground state of the system. However, the excess
of one sort of nucleons blocks certain
$\nu$-cells from participating in alpha-clustering.
To describe these blocked states we define the effective occupation numbers
$f_{\nu}$ in such way that $f_{\nu}=1$ if the set $\nu$ is blocked (partially
occupied) and $f_{\nu}=0$ if the set $\nu$ is available for alpha scattering.
Using these occupation factors we have to introduce the new sums $\sum_{\nu}{}^{'}$
which include only available sets of single-particle orbits
\begin{equation}
\sum_{\nu} \rightarrow \sum_{\nu}{}^{'}=\sum_{\nu}(1-f_\nu). \label{79}
\end{equation}
The excess of the nucleons of one sort plays effectively the same role in this problem
as the increase of temperature in a situation of standard pairing, where the blocking factor
$f_{\nu}$ would be proportional to $\exp(-E_{\nu}/T)$, where $E_{\nu}$ is the energy
of an excitation (an unpaired particle or a broken pair) responsible for blocking.

The presence of blocked states clearly weakens the alpha condensate and pushes up
the ground state energy (\ref{78}). Although, differently from equilibrium thermal excitation,
there are many ways of distributing the extra particles over single-particle orbits,
to estimate this effect we suggest a smooth behavior of the occupation numbers
$f_{\nu}$ as function of $\epsilon_{\nu}$. Thus, replacing the sum (\ref{79}) by
an integral, we can write
\begin{equation}
\int \rho(\epsilon) d\epsilon \rightarrow \int \bar{\rho}(\epsilon) d\epsilon \simeq
\bar{\rho}_0 \int d\epsilon,                               \label{80}
\end{equation}
where $\bar{\rho}_0=\rho_0-\delta \rho$ is the reduced
density of pair states near
the Fermi surface and $\delta \rho \propto \rho_0 |N-Z|$ is the density of blocked states.
Assuming this, we can conclude that moving away from the line $N=Z$ we decrease the
alpha-correlation energy in eq. (\ref{78}) mainly because of quenching
the alpha-gap $\Delta^{(\alpha)}\sim \exp(-2/C \bar{\rho}_0)$.

In contrast to the alpha condensate, the usual pairing is not affected by a different
number of protons and neutrons. Indeed, at non-zero values of $|N-Z|$, an even-even
system simply splits into two superfluid subsystems, proton and neutron, and each subsystem
uses all available space without feeling any blocking. At some value of $|N-Z|$ it
will be not favorable anymore to support alpha-condensate and the system will change
its internal structure from alpha-clusters to pairs. The moment when it happens
as a first order phase transition can be simply estimated as a crossing point
of equal values for pairing gaps and alpha-gap: $\bdel \sim \Delta^{(\alpha)}$.
This leads to the following condition,
\begin{equation}
\frac{1}{2}\,\bar{\rho}_0 C = \rho_0 G.                      \label{81}
\end{equation}
This condition makes the constants in front of the large logarithm, see eqs.
(\ref{51},\ref{52}), equal to each other that corresponds to the crossing
of ground state energies.

\section{Conclusion}

In this article we considered a qualitative behavior of a two-component
(``neutron-proton") fermionic matter with the attractive interactions of two
types, pairing on time-conjugate orbits and quartic (``alpha") correlations.
We employed a new trial wave function that extends a classical
BCS ansatz. The variational procedure leads to the set of equations for
the occupation factors of the ground state in each cell of four single-particle
states. Similarly to the BCS method for large systems, we constructed a
logarithmic approximation for solving these equations. The main result is
that, keeping only the large logarithms, we see the absence of coexistence
of two possible condensates.
This result is strictly valid for the system
with equal populations of two kind of fermions ($N=Z$ in the nuclear case).

Assuming that the ground state of a symmetric system corresponds to the alpha
condensate we propose that the growing excess of the fermion number of one kind
will lead to the first order phase transition to the standard case of two
fermionic condensates. The analog of this consideration for finite nuclei
can be sought for in the vicinity of the doubly magic nucleus $^{100}$Sn.
The significant deviations of electromagnetic transition probabilities
from typical shell-model regularities, see for example \cite{starosta07},
might be related to the presence of alpha-correlations beyond $^{100}$Sn
up to some critical value of the neutron excess. Of course, in realistic
nuclear applications the logarithmic approximation will not be sufficient and
it will be necessary to perform detailed specific calculations with
appropriate nuclear orbitals and interaction parameters. Another area of
possible applications can be related to the crust of neutron stars.

One comment might be appropriate here. The alpha correlations considered above
are different from the bosonic alpha condensate intensely discussed recently
\cite{tohsaki01} in relation to cluster levels in light nuclei like $^{12}$C
and $^{16}$0. The interrelation here is similar to that between the BCS
theory and an older Blatt-Butler-Schafroth theory of superconductivity
\cite{blatt55} based on the idea of a dilute gas of bosonic quasimolecules
made of electron pairs.

R.S. is grateful to A. Pomeransky for useful discussions. The support from the DOE grant DF-FC02-09ER41584 and NSF grant PHY-0758099 is acknowledged.

\end{document}